\documentclass[aps,nofootinbib]{revtex4}
\usepackage{amstext,amsmath,amssymb,amsfonts,bbm}
\usepackage[latin1]{inputenc}
\usepackage{fancyhdr}
\usepackage[dvips]{graphicx}
\usepackage{epsfig}

\def\w{\wedge} 
\def\la{\langle} \def\ra{\rangle}
\def\f{\frac}
\def\om{\omega}
\newcommand{\SU}{\mathrm{SU}}
\newcommand{\SO}{\mathrm{SO}}
\newcommand{\ISO}{\mathrm{ISO}}

\newcommand{\lalg}[1]{\mathfrak{#1}}
\newcommand{\su}{\lalg{su}} \renewcommand{\sl}{\lalg{sl}}
 
\newcommand{\so}{\lalg{so}} 
 
\newcommand{\lieg}{\lalg{g}}
 \def\dag{^\dagger}
\def\tl{\widetilde}
\def\pp{\partial}
\def\nn{\nonumber} \def\arr{\rightarrow}
\def\eps{\epsilon}
\def\what{\widehat}
\newcommand{\tr}{\mathrm{tr}}
\def\dag{^\dagger}
\def\C{{\mathbbm C}}

\def\be{ \begin{equation}}
\def\ee{\end{equation}}
\def\bes{\begin{eqnarray}}
\def\ees{\end{eqnarray}}

\newcommand{\Ref}[1]{(\ref{#1})}

\begin{document}
\title{{\large A Immirzi-like parameter for 3d quantum gravity}}
\author{{\bf Valentin Bonzom}\email{valentin.bonzom@ens-lyon.fr}}
\author{{\bf Etera R. Livine}\email{etera.livine@ens-lyon.fr}}
\affiliation{Laboratoire de Physique, ENS Lyon, CNRS UMR 5672, 46 All\'ee d'Italie, 69364 Lyon Cedex 07, France}

\begin{abstract}
We study an Immirzi-like ambiguity in three-dimensional quantum gravity. It shares some features
with the Immirzi parameter of four-dimensional loop quantum gravity: it does not affect the
equations of motion, but modifies the Poisson brackets and the constraint algebra at the canonical
level. We focus on the length operator and show how to define it through non-commuting fluxes. We
compute its spectrum and show the effect of this Immirzi-like ambiguity. Finally, we extend these
considerations to 4d gravity and show how the different topological modifications of the action
affect the canonical structure of loop quantum gravity.

\end{abstract}
\maketitle

\section*{Introduction}

Loop quantum gravity (LQG) presents a framework for a canonical quantization of general relativity
(see e.g \cite{carlo}). It defines a Hilbert space of quantum states of (space) geometry, spanned
by the spin network states, and constraint operators implementing the invariance of the theory
under space-time diffeomorphisms. It then derives discrete spectra for geometric quantities such as
the areas and volumes. This whole framework is nevertheless affected by a quantization ambiguity,
parametrized by the Immirzi parameter $\gamma$ \cite{immirzi}. It is the parameter for a canonical
transformation on the phase space, it does not have any effect on the classical equation of motion
but it translates into a non-unitary transformation at the quantum level thus leading to
non-equivalent quantization
\cite{immirzi,immirzi2}. It scales the spectra of geometric operators (the length unit gets a
$\sqrt{\gamma}$ factor) but also affects the Hamiltonian constraint.

At the level of the classical action, Holst showed that the Immirzi parameter is introduced by
adding a new term to the first order Palatini action for general relativity (GR) \cite{holst}. This
new term is the square of the torsion up to the Nieh-Yan topological invariant. It does not change
the equations of motion (as long as the metric is non-degenerate) which remain equivalent to the
Einstein equations. Furthermore, in the McDowell-Mansouri reformulation of GR as a constrained BF
theory for a $\SO(4,1)$ gauge group, $\gamma$ amounts to introducing a quadratic potential for the
Lagrange multiplier $B$ \cite{McD}. Physically, it has also been shown that the Immirzi parameter
is related to CP violations for fermion fields \cite{coupling to fermions}. Finally, since the
Immirzi parameter is due to an extra term in the classical action, it will affect the path integral
even though it does not change the field equations. There has been recent proposals to take it into
account in spin foam models for the LQG path integral \cite{spinfoam}.

In the present work, we study a similar ambiguity in 3d quantum gravity. From the point of view of
the generalized Palatini action introduced by Holst \cite{holst}, the Immirzi ambiguity appears
because of the existence of two non-degenerate invariant bilinear forms on the Lorentz algebra
$\so(3,1)$ (or $\so(4)$ in the Riemannian case). The same idea is applied in three dimensions and
again leads to an ambiguity in the action, as first noted by Witten in \cite{witten:88}. Gravity
with a cosmological constant in three space-time dimensions is a topological BF field theory. We
can then introduce an extra Chern-Simons-like term which does not modify the equations of motion
for pure gravity. This term affects the path integral \cite{witten:88}. However, although 3d loop
quantum gravity has been thoroughly studied \cite{tout,3d lqg,thomas3d}, this Immirzi-like
ambiguity has not been investigated in details at the canonical level as far as we know (it was
nevertheless briefly mentioned in \cite{tout} as the $\theta$-ambiguity). The natural questions is
whether or not it leads to inequivalent loop quantizations like in the 4d theory and how does it
affect the geometrical operators at the kinematical level.

We first review in section \ref{review 3d} some basic facts about 3d gravity reformulated as a
Chern-Simons action for an extended $\so(3,1)$-connection. We then introduce the Immirzi parameter
for 3d gravity in section \ref{immirzi hamiltonian}. We perform the Hamiltonian analysis, carefully
looking at the constraints algebra. The quantization is studied in section \ref{quantization},
where we have to deal with non-commuting fluxes. We focus on the construction of the length
operator. The Immirzi parameter alters its spectrum by a shift instead of the usual scaling derived
in 4d loop quantum gravity. Moreover it gives rise to a new ambiguity: the curve whose length is
measured has to be labeled with a $\su(2)$-representation.
Finally, we extend these considerations to the 4d case in section \ref{comparison} and discuss the
effect on the theory's canonical structure of the various topological terms that we can add to the
action.

\section{\label{review 3d} A quick review of 3d gravity}

Let us consider a 3-manifold $M$ and a principal $G$-bundle over $M$. We focus on $G=\SU(2)$ for 3d
Riemannian gravity. We call $\mathfrak{g}$ the Lie algebra of $G$. Gravity is formulated in term of
a triad field $e$ and the spin connection $\om$. $e$ is a $\mathfrak{g}$-valued one form and $\om$
is a $\mathfrak{g}$-connection form whose curvature is denoted by $F[\om]$. Internal indices are
contracted with the Killing metric $\delta_{ij}$ on $\mathfrak{g}$ ($i,j=1,2,3$). The action of 3d
gravity with cosmological constant $\Lambda$ reads:
\begin{equation} \label{gravity action}
S_{GR}(e,\omega)=\,\f1{4\pi G}\int \left[2\,e^i\wedge
F_i[\om]\,+\,\frac{\Lambda}{3}\epsilon_{ijk}e^i\wedge e^j\wedge e^k\right].
\end{equation}
In the following, we will systematically forget Newton's constant and set $4\pi G\equiv 1$. The
equations of motion\footnotemark impose a vanishing torsion, $d_\omega e^i=0$, and a constant
curvature given by the cosmological constant, $F^i+(\Lambda/2)\epsilon^i_{\phantom{i}jk}e^j\wedge
e^k=0$.
The action \eqref{gravity action} is invariant under $\SU(2)$ gauge transformations:
$$
 e\,\arr\, geg^{-1},\qquad \om\,\arr\,g\om g^{-1}+gd g^{-1},
$$
with $g\in\SU(2)$. It is also invariant under a translational symmetry:
\begin{equation}
\delta\omega^i=\Lambda \epsilon^i_{\phantom{i}jk}\ e^j\chi^k \qquad\mathrm{and} \qquad\delta e^i=d_\omega \chi^i.
\end{equation}
This symmetry is also called 'topological' symmetry since it implies that the field $e$ is pure
gauge and is responsible for the lack of local degrees of freedom of the theory. Space-time
diffeomorphisms can be generated as a combination of both types of transformations.

\footnotetext{The covariant derivative is defined in term of the connection as $(d_\omega
v)^i=dv^i+[\omega,v]^i=dv^i+\epsilon^i_{\phantom{i}jk}\omega^j\wedge v^k$.}

These two symmetries can be unified into a single gauge symmetry by enlarging the rotation group
$G$ to a larger gauge group $\tl{G}$. This idea is the key of the reformulation of 3d gravity as a
Chern-Simons theory for $\tl{G}$ \cite{witten:88}. The larger gauge group $\tl{G}$ is $\SO(4)$,
$\ISO(3)$ or $\SO(3,1)$ depending on whether $\Lambda$ is positive, zero or negative. The
generators of the Lie algebra $\tl{\lieg}$ satisfy the commutation relations:
\begin{equation}
[J_i,J_j]=\epsilon_{ij}^{\phantom{ij}k} J_k, \quad\quad [J_i,K_j]=\epsilon_{ij}^{\phantom{ij}k}
K_k, \quad\quad [K_i,K_j]=s\ \epsilon_{ij}^{\phantom{ij}k} J_k,
\end{equation}
where $s=-1,0,1$ is the sign of the cosmological constant. We now define a connection for the
enlarged gauge group, $A=\omega^iJ_i+\sqrt{\lvert\Lambda\rvert}\ e^iK_i$. A rotation generated by
an element $u=u^iJ_i$ gives a $\SU(2)$ gauge transformation, while boosts $v=v^iK_i$ give the
translational symmetry. To build a Chern-Simons action, we choose a non-degenerate invariant
bilinear form on the Lie algebra $\tl{\lieg}$:
\begin{equation} \label{forme killing}
\langle J_i,K_j\rangle=\delta_{ij} \qquad\qquad \langle J_i,J_j\rangle=\langle K_i,K_j\rangle=0.
\end{equation}
Then, for a non-vanishing cosmologic constant $\Lambda\ne0$, the action for 3d gravity can be
written as a Chern-Simons theory:
\begin{equation} \label{reformulation witten}
S(A)=S_{GR}(\omega,e)=\frac{1}{\sqrt{\lvert\Lambda\rvert}}\int_M d^3x\ \epsilon^{\mu\nu\rho}\big(
\langle A_\mu,\partial_\nu A_\rho\rangle+\frac{1}{3}\langle A_\mu,[A_\nu,A_\rho]\rangle\big)
\end{equation}
The case of a vanishing cosmological constant is recovered by setting $s=0$ and
$\sqrt{\lvert\Lambda\rvert}=1$ (or to any other arbitrary constant). The equations of motion simply
say that the curvature $R$ of the connection $A$ vanishes. Since $R=\big(F^i+\frac{\Lambda}{2}\
[e,e]^i\big)\ J_i+\big(d_\omega e\big)^i\ K_i$, this is equivalent to the previous equations of
motion with a vanishing torsion and a constant curvature.

\section{\label{immirzi hamiltonian} The Immirzi ambiguity at the classical level}

\subsection{Generalizing the action}


When $\Lambda\neq 0$, there exists another invariant non-degenerate bilinear form on $\tl{\lieg}$.
It is related to the one used above by the Hodge operator $\star$ exchanging the rotations $J_i$
with the boosts $K_i$: $(B,C)=\langle B,\star C\rangle$, with $(\star J)_{IJ}\,\equiv\,
\epsilon^{IJ}{}_{KL} J^{KL}/2$. It is given explicitely by
\begin{equation} \label{second bilinear form}
(J_i,J_j)=\delta_{ij},\qquad (K_i,K_j)=s\ \delta_{ij}\quad\mathrm{and}\quad(J_i,K_j)=0.
\end{equation}
We define the associated Chern-Simons action :
\begin{equation} \label{second action}
\tl{S}(A)
=\frac{1}{\sqrt{\lvert\Lambda\rvert}}\int_M d^3x\ \epsilon^{\mu\nu\rho}\big(  (A_\mu,\partial_\nu
A_\rho)+\frac{1}{3}( A_\mu,[A_\nu,A_\rho])\big).
\end{equation}
Let us write it in $(e,\omega)$ variables :
\bes \label{second action bis}
\tl{S}(A) &
=&\frac{1}{\sqrt{\lvert\Lambda\rvert}}\int_M \,\omega^i\wedge d\omega_i+
\frac{1}{3}\epsilon_{ijk} \omega^i\wedge \omega^j\wedge\omega^k+s\lvert\Lambda\rvert\, e^i\wedge d_\omega e_i \nn\\
&=&\frac{1}{\sqrt{\lvert\Lambda\rvert}}S_{CS}(\omega)+s\sqrt{\lvert\Lambda\rvert}\ \int_M
\, e^i\wedge d_\omega e_i.
\ees
It is straightforward to check that this action actually gives the same equations of motion as the
gravity action \eqref{gravity action} for $s\neq 0$.
We now consider a linear combination of both actions, $S_\gamma(A)=S(A)+\gamma^{-1}\tilde{S}(A)$.
The equations of motion are again equivalent to the Einstein equations~:
\begin{equation}
\left\{\begin{array}{lr}
\big(F^i+s\ \frac{\lvert\Lambda\rvert}{2}\,
\epsilon^{i}_{\phantom{i}{jk}}e^j\wedge e^k\big) +s\ \frac{\sqrt{\lvert\Lambda\rvert}}{\gamma}\ d_\omega e^i&=0 \\
d_\omega e^i+ \frac{1}{\gamma\sqrt{\lvert\Lambda\rvert}}\big(F^i+s\ \frac{\lvert\Lambda\rvert}{2}\
\epsilon^{i}_{\phantom{i}{jk}}e^j\wedge e^k\big) &=0
\end{array} \right.
\end{equation}
Providing that $\gamma^2\neq s$, we get a one-parameter family of theories classically describing
3d gravity\footnotemark. We call $\gamma$ the Immirzi parameter in analogy with the parameter
entering the Holst action in 4d since they both appear through Hodge duality.

The case of a vanishing cosmological constant $\Lambda=0$ works in a similar way as above. We can
derive by formally setting $s=0$ and $\lvert\Lambda\rvert=1$. Then the second bilinear form
\Ref{second bilinear form} becomes degenerate and the new action $\tl{S}(A)$ reduces to the
Chern-Simons action $S_{CS}(\om)$ for the spin connection.

\footnotetext{
There are all classically equivalent to pure gravity. The coupling to matter fields will depend on
$\gamma$ (see in appendix for more details). This is the same situation as in 3+1 dimensions where
the Immirzi parameter affects the effective dynamics of fermions \cite{coupling to fermions}.}

The particular choices $\gamma^2=s$ correspond to restricting the Chern-Simons connection to its
self-dual or anti-self-dual component. Indeed, we can use the Hodge operator to split the
connection int two: $A=A_++A_-$ with $\star A_\pm=\pm\sigma A_\pm$. We have
$A^i_\pm=\omega^i\pm\sigma\sqrt{\lvert\Lambda\rvert}e^i$ where $\sigma^2=s$, explicitly $\sigma=1$
for $\Lambda>0$ and $\sigma=i$ for $\Lambda<0$. Then the full action also splits in two:
\begin{equation} \label{self-dual decomposition}
S_\gamma(A)\,=\,\left(\gamma^{-1}+s\sigma\right)\, S_{CS}(A_+)\,+\,
\left(\gamma^{-1}-s\sigma\right)\, S_{CS}(A_-)
\end{equation}
When $\gamma$ is infinite, $\gamma^{-1}=0$, $S_\gamma(A)$ is the original action $S(A)$ and the two
parts of the action $S_\pm$ have opposite coupling constants. This relation is responsible for the
link between the Turaev-Viro invariant and the Reshetikhin-Turaev invariant \cite{using quantum
groups}. For finite values of the Immirzi parameter, this no longer holds.
%
In the following, we will always work at $\gamma^2\ne s$.

\subsection{Canonical analysis}

To perform the Hamiltonian analysis, we assume that $M$ is of the form $M=\Sigma\times\mathbb{R}$,
where $\Sigma$ is a two-dimensional smooth manifold of arbitrary topology. We choose arbitrary
coordinates $x^a=(x^1,x^2)$ on the canonical surface $\Sigma$ and complete it with a coordinate
time $x^0$ on $\mathbb{R}$. Following this 2+1 splitting, we write the action as :
\begin{equation} \label{canonical decomposition} \begin{split}
S_\gamma& =
\int d^3x\ 2\epsilon^{ab}\delta_{ij}(e_b^i\ \partial_0\omega_a^j+
\frac{1}{2\gamma\sqrt{\lvert\Lambda\rvert}}\ \omega_b^i\ \partial_0\omega_a^j
+s\ \frac{\sqrt{\lvert\Lambda\rvert}}{2\gamma}\ e_b^i\ \partial_0 e_a^j) \\
&+2\epsilon^{ab}\delta_{ij}(\omega_0^j+s\ \frac{\sqrt{\lvert\Lambda\rvert}}{\gamma}e_0^j)\ D_a
e_b^i +\epsilon^{ab}\delta_{ij}(e_0^i+\frac{1}{\gamma\sqrt{\lvert\Lambda\rvert}}\omega_0^i)\
(F_{ab}^j+\Lambda\epsilon^{j}_{\phantom{j}kl}e_a^k e_b^l)
\end{split} \end{equation}

The kinematical terms of \eqref{canonical decomposition} involving time derivative $\pp_0$
determine the Poisson brackets.
The symplectic structure explicitly depends on the parameter $\gamma$:
\begin{align} \label{we bracket}
\{\omega_a^i(x),e_b^j(y)\}& =\frac{1}{2}\frac{\gamma^2}{\gamma^2-s}\ \epsilon_{ab}\delta^{ij}\delta^{(2)}(x-y), \\
\label{ww bracket}
\{\omega_a^i(x),\omega_b^j(y)\}& =\frac{\sqrt{\lvert\Lambda\rvert}}{2}\frac{s\gamma}{s-\gamma^2}\ \epsilon_{ab}\delta^{ij}\delta^{(2)}(x-y), \\
\label{ee bracket}
\{e_a^i(x),e_b^j(y)\}& =\frac{1}{2\sqrt{\lvert\Lambda\rvert}}\frac{\gamma}{s-\gamma^2}\ \epsilon_{ab}\delta^{ij}\delta^{(2)}(x-y)
\end{align}
The connection $\om$ is still conjugate to the triad field $e$. But now both fields $e$ and $\om$
have become non-commutative. We recover the usual canonical structure,
$\{\om,e\}=\eps\delta,\,\{\om,\om\}=0=\{e,e\}$, in the limit $\gamma\arr\infty$. On the other hand,
to get the symplectic structure for $\Lambda=0$, we need to set $\lvert\Lambda\rvert=1$ and $s=0$.
Then the connection $\om$ becomes commutative while the triad $e$ remains non-commutative.

Notice that it is always possible to find two constants $\alpha$ and $\beta$ such that the fields
$\omega_a^i+\alpha e_a^i$ and $\omega_a^i+\beta e_a^i$ are both commutative and canonically
conjugate to each other. However both fields are connections on $\Sigma$, and hence the loop
quantization based on a connection and its conjugate vierbein can not be applied straightforwardly.

The two remaining terms in the action \eqref{canonical decomposition} give the hamiltonian. It is
simply a linear combination of constraints imposed by the Lagrange multipliers $e_0^i$ and
$\om_0^i$:
\begin{equation}
\epsilon^{ab}D_a e_b^i=0\quad\mathrm{and}\quad F_{ab}^i+s\lvert\Lambda\rvert\epsilon^i_{\phantom{i}jk}e_a^j
e_b^k=0.
\end{equation}
These constraints are the same as in usual gravity. They are first class and represent the Lie
algebra $\tl{\lieg}$. We define the smeared constraints:
\begin{align}
G(\lambda) &=2\frac{\gamma^2-s}{\gamma^2}\int_\Sigma d^2x\ \epsilon^{ab}\delta_{ij}\ \lambda^i(x)\ D_a e_b^j(x), \\
H(\lambda) &=\frac{\gamma^2-s}{\gamma^2\sqrt{\lvert\Lambda\rvert}}\int_\Sigma d^2x\,
\epsilon^{ab}\delta_{ij}\ \lambda^i(x)\ \big(F_{ab}^j(x)+s\lvert\Lambda\rvert\
\epsilon^i_{\phantom{i}jk}\ e_a^j(x) e_b^k(x)\big).
\end{align}
In the limit $\gamma\arr\infty$, $G(\lambda)$ generates infinitesimal $\SO(3)$ transformation
corresponding to gauge transformations on $A$ with parameter $\lambda^iJ_i$ while $H(\lambda)$
generates the translations corresponding to gauge transformations on $A$ with parameter
$\lambda^iK_i$. Since the Immirzi parameter affects the Poisson bracket, it also modifies the
commutation relation of the constraints $G(\lambda)$ and $H(\lambda)$. This results in a change of
representation basis for the Lie algebra $\tilde{\mathfrak{g}}$. Indeed $G(\lambda)$ now induces
the infinitesimal transformation defined by $\lambda^i(J_i-\gamma^{-1}K_i)$ :
\begin{align}
\big\{G(\lambda),\omega_a^i(x)\big\}& =D_a\lambda^i(x)-\gamma^{-1}\epsilon^i_{\phantom{i}jk} \sqrt{\lvert\Lambda\rvert}\ e_a^j(x)\lambda^k(x), \\
\big\{G(\lambda),\sqrt{\lvert\Lambda\rvert}\ e_a^i(x)\big\}& =\epsilon^i_{\phantom{i}jk}\ \sqrt{\lvert\Lambda\rvert}\ e_a^j(x)\lambda^k(x)-\gamma^{-1}D_a\lambda^i(x),
\end{align}
while $H(\kappa)$ is related to the gauge transformation $\kappa^i(K_i-s\gamma^{-1}J_i)$ :
\begin{align}
\big\{H(\kappa),\omega_a^i(x)\big\}& =\epsilon^i_{\phantom{i}jk} \sqrt{\lvert\Lambda\rvert}\ e_a^j(x)\kappa^k(x)-s\gamma^{-1}\ D_a\kappa^i(x), \\
\big\{H(\kappa),\sqrt{\lvert\Lambda\rvert}\ e_a^i(x)\big\}& =D_a\kappa^i(x)-s\gamma^{-1}\epsilon^i_{\phantom{i}jk} \sqrt{\lvert\Lambda\rvert}\ e_a^j(x)\kappa^k(x).
\end{align}
Then the constraints algebra naturally reflects the commutation relations of $\tilde{\mathfrak{g}}$
in the basis $(J_i-\gamma^{-1}K_i,\ K_i-s\gamma^{-1}J_i)$~:
\begin{align}
\{ H(\kappa),H(\lambda)\}& =s\ G([\kappa,\lambda])-s\gamma^{-1}\ H([\kappa,\lambda])  \\
\{ G(\lambda),H(\kappa)\}& =H([\lambda,\kappa])-s\gamma^{-1}\ G([\lambda,\kappa])  \\
\{ G(\lambda),G(\kappa)\}& =G([\lambda,\kappa])-\gamma^{-1}\ H([\lambda,\kappa]).
\end{align}
It is clear that we can find two linear combinations of the constraints which act as the usual
rotation and translation constraints. Therefore we expect that the physical Hilbert space of the
quantum theory
will not changed. Nevertheless, observables and the action of the corresponding quantum operators
will be modified by the Immirzi parameter. In the next section, we study the length operator in the
context of a loop quantization.

\subsection{Comparing with 3d Yang-Mills theory}

In the limit case $\Lambda=0$, the Poisson bracket simplifies and leads to the phase space of a
system with non-commutative momenta:
\begin{align}
\label{conn. brackets}
\{\omega_a^i(x),\omega_b^j(y)\}& =0, \\
\{\omega_a^i(x),e_b^j(y)\}& =\frac{1}{2}\ \epsilon_{ab}\delta^{ij}\delta^{(2)}(x-y), \\
\label{triad brackets}
\{e_a^i(x),e_b^j(y)\}& =-\frac{1}{2\gamma}\ \epsilon_{ab}\delta^{ij}\delta^{(2)}(x-y).
\end{align}
This can be read directly from the action:
\be
S_\gamma\,=\, 2\int e^i\w F_i[\om] + \f1\gamma S_{CS}(\om)
\,=\,
2\int \left(e^i+\f1{2\gamma}\om^i\right)\w dw_i + \left(e+\f1{3\gamma}\om\right)\w \om \w \om.
\ee
Notice that we can not re-absorb entirely the Chern-Simons term in a redefinition of the triad
field $e$. The canonical variables are now $\omega_a^i$ and
$\pi^a_i=\eps^{ab}(e_{bi}+\f{1}{2\gamma}\omega_{bi})$. Let us insist on the fact that the momentum
$\pi$ does not transform as the triad $e$ under $\SU(2)$-gauge transformation but as a connection.
The translational symmetry is still generated by the constraint $H(\lambda)=0$ while $\SU(2)$ gauge
transformed are now generated by the constraint $G(\lambda)+\f{1}{\gamma}\ H(\lambda)=0$. Notice
that the local form of the latter is: $\eps^{ab}\ (D_a e_b^i+\f{1}{2\gamma}F_{ab}^i)=D_a
\pi^a_i+\f{1}{2\gamma}\eps^{ab}\pp_a\omega_{bi}=0$.

A simple example of a situation involving non-commutative momenta is the Landau problem of a
particle in a magnetic field (see e.g. \cite{landau}). Because the Lagrangian couples the velocity
of the particle to the vector potential, the canonical momentum is the velocity shifted by the
vector potential. Thus the brackets between velocity components do not vanish and are proportional
to the magnetic field. This well-known situation has a field-theoritical analog (see e.g.
\cite{TMYM}) which is of importance for us : the 3d Yang-Mills theory with an additional Chern-Simons term. This
theory, known as topologically massive gauge theory \cite{TMYM}, yields a massive gauge boson. The
interesting point for us is that the phase space is the same as in 3d gravity, the mass of the
gauge boson being just the (inverse of the) Immirzi parameter.

Indeed, let us consider, without loss of generality, the following abelian action:
\be
S_{TMYM}(A)=-\f{1}{4}\int d^3x\ F_{\mu\nu}F^{\mu\nu}+\f{m}{2}\int d^3x\ \eps^{\mu\nu\lambda}\
A_\mu\pp_\nu A_\lambda.
\ee
$A_\mu$ is a gauge field (of mass dimension $1/2$) and $m$ is the mass of the
photon\footnote{First, one can check that the field equation for the vector
$G_\mu=\eps_{\mu\nu\sigma}F^{\nu\sigma}$ is that of a free massive field:
$$
(\pp^2+m^2)G_\mu=0.
$$
At the quantum level, the Feynman propagator has a singularity at $p^2=m^2$, showing that the
photon has become massive. Explicitly, we get in the Landau gauge:
$$
D_{\mu\nu}=\f{1}{p^2-m^2}\big(-\eta_{\mu\nu}+\f{p_\mu
p_\nu}{p^2}+i\f{m}{p^2}\eps_{\mu\nu\sigma}p^\sigma\big)
$$
The representation theory of the (2+1)d-Poincaré algebra shows that the photon has one polarization
per momentum. We refer to \cite{TMYM} for more details.}. From the Hamiltonian point of view, when
$m=0$, the momentum conjugated to the spatial connection $A_a$ ($a=1,2$) is as usual the electric
field $E_a$. For $m\neq 0$, the momentum becomes $\Pi_a=E_a+\f{m}{2}\eps_{ab}\ A_b$ and the Poisson
brackets for the electric field is the same as those of the triad components with zero cosmological
constant \eqref{triad brackets}:
\be
\{E_a(x),E_b(y)\}=m\ \eps_{ab}\ \delta^{(2)}(x-y),
\ee
with the identification $m=-1/(2\gamma)$. The form of the Hamiltonian is unchanged,
$H=\f{1}{2}(E^2+B^2)$. Thus, the Chern-Simons term does not modify the classical energy but alters
the relations between momenta and electric fields. In gravity, the Chern-Simons term preserves the
gauge symmetries -- in particular, the theory is still topological -- but modifies the expressions
of their generators.

The Gauss constraint, which generates the gauge transformations, reads $\pp_a
E_a+m\eps_{ab}F_{ab}=\pp_a \Pi_a+m\eps_{ab}\pp_a A_b=0$. In the usual quantization, $\Pi$ (resp.
$\pi$ for 3d gravity) becomes a functional derivative with respect to the connection. But $\Pi$
(resp. $\pi$) is a connection while a functional derivative should transform covariantly under a
gauge transformation. Then the quantization of the Gauss constraint implies to work with states
which are not exactly gauge invariant \cite{Hamiltonian TMYM}. When dealing with a background
independent theory such as 3d gravity, the loop quantization avoids those difficulties and provides
a well-defined gauge invariant and diff-invariant state space.

Leaving aside the topological constraints, we now proceed to the quantization of 3d gravity with
the Immirzi parameter at the kinematical level, by quantizing the algebra of loop variables.

\section{\label{quantization} Quantization of lengths with the Immirzi parameter}

The quantization of 3d gravity has been extensively studied. In particular, it was one of the first
example of the loop quantization programme \cite{tout,thomas3d} and has since been completed (see
e.g. \cite{3d lqg}, see also \cite{using quantum groups,using CS} for quantum groups and/or
Chern-Simons approaches). We first work with $\Lambda=0$ and define modified flux observables which
enable us to compute the length spectrum. This will generalize straightforwardly to the generic
$\Lambda\ne 0$ case.

\subsection{Quantizing the loop variables}

As we have seen above, the Poisson brackets for $\Lambda=0$ reduce to:
\bes
\{\omega_a^i(x),\omega_b^j(y)\}& =&0, \nn\\
\{\omega_a^i(x),e_b^j(y)\}& =&\frac{1}{2}\ \epsilon_{ab}\delta^{ij}\delta^{(2)}(x-y), \label{p1}\\
\{e_a^i(x),e_b^j(y)\}& =&-\frac{1}{2\gamma}\ \epsilon_{ab}\delta^{ij}\delta^{(2)}(x-y). \nn
\ees
Following the standard loop quantization, we would like to quantize the holonomy-flux algebra (see
e.g. \cite{thomaslqg}). Since the brackets between $\om$'s components vanish, we build the
kinematical Hilbert space as usual: we use gauge invariant wavefunctionals of the connection which
depend only on the holonomies of $\om$ along the edges of some finite graph embedded in the
canonical surface $\Sigma$. A basis of this Hilbert space of quantum geometry states is provided by
the spin network states. The difficulty now resides in the non-commutativity of the triad
components. We need to quantize the fluxes of the triad field so that the quantum commutation
relations reflect the non-vanishing Poisson brackets of $e$'s components. We could first try adding
to the usual action of the flux a term which acts by multiplication. The latter usually acts on
holonomies by Poisson brackets :
\begin{equation}
\left.\what{\int_c e^i}\,\right|_{\gamma=+\infty}\,=\,-i\left\{\int_c e^i,\,\cdot\right\}
\end{equation}
for a curve $c$. In order to faithfully represent the Poisson brackets \Ref{p1}, the right term to
add would be a simple function of the connection proportional to $\gamma^{-1}\int_c \omega^i$.
However, this choice is not consistent with the different transformation properties of the
connection and triad fields.

Our solution is to consider other observables, such that the new term to add to the triad operators
is simply the holonomy of $\omega$ along the curve $c$, whose action is well defined by
multiplication on spin-network states. For a curve $c$ going from $x_0$ to $x_1$ we call
$U^{x_0}_{x_1}\in\SU(2)$ (or $U(c)$) the holonomy of $\om$ along $c$  and we write
$V^{x_1}_{x_0}\equiv U^{x_1}_{x_0}$ (or $V(c)$) for its inverse. We also assign a $\SU(2)$
representation of spin $j$ to the curve $c$ and we denote the $\su(2)$ generators
$\tau_i=-i\sigma_i/2$ where $(\sigma_i)_{i=1,2,3}$ are the usual Pauli matrices.
Following \cite{cattaneo}, we introduce the matrix $O(c)$ in the spin $j$ representation, whose
matrix elements are :
\begin{equation}
O(c)^\alpha_\beta=\int_c \big(\ V^y_{x_0}\ e_{|y}\ V^{x_1}_y\ \big)^\alpha_\beta,
\end{equation}
where the holonomies are taken in the spin $j$ representation and $e_{|y}=e^i_a(y) dx^a
\tau^{(j)}_i$.
This observable $O(c)$ amounts to inserting the triad in the holonomy and integrate over the
position of the insertion. Notice that when $c$ is a closed curve, the quantity $\tr_{(j)} (O(c))$
is actually the one-insertion loop variable as used in the original formulation of loop
quantization \cite{tout}. It is $\SU(2)$ gauge invariant and is also invariant under infinitesimal
topological transformations, $\delta\omega=0$ and $\delta e=d_\omega \chi$, which do not move the
basepoint $x_0$ of the loop $c$, $\chi(x_0)=0$. We also introduce the adjoint matrix $O(c)\dag$:
\be
O^\dagger(c)=-\int_c U^y_{x_1}\ e_{|y}\ U^{x_0}_y.
\ee
We first describe the action of $O(c)$ and $O^\dagger(c)$ at the quantum level when there is no
Immirzi parameter $\gamma\arr\infty$. They become operator-valued matrices, by replacing the triad
field $e$ by a functional derivative with respect to the connection. Since they consist in a single
triad insertion, they act on holonomies locally.
Assuming that the curve $c$ crosses the graph of the spin-network graph $\Gamma$ just once, at a
point $p$ belonging to the link $\ell$, then the action of $\what{O}(c)$ and $\what{O}^\dagger(c)$
on the spin network state is given by their action on the matrix elements  of the holonomy
$U(\ell)^a_b$ along the link $\ell$ (taken in the spin $j_\ell$ representation), that is:
\begin{align} \label{crochet O}
-i\{O(c)^\alpha_\beta,U(\ell)^a_b\} & =-\frac{i}{2}\ \epsilon(c,\ell)\ \delta^{ik}\ \Big(V^p_{x_0}\
\tau^{(j_c)}_i\ V^{x_1}_p\Big)^\alpha_\beta\, \Big(U^A_p\, \tau^{(j_\ell)}_k\,
U_B^p\Big)^a_b  \\ \label{crochet Odagger}
-i\{O(c)^{\dagger\alpha}_{\phantom{\dagger}\beta},U(\ell)^a_b\} &=\frac{i}{2}\ \epsilon(c,\ell)\
\delta^{ik}\ \Big(U^p_{x_1}\ \tau^{(j_c)}_i\ U^{x_0}_p\Big)^\alpha_\beta\, \Big(U^A_p\,
\tau^{(j_\ell)}_k\, U_B^p\Big)^a_b
\end{align}
where $A,B$ are respectively the start and target vertex of the link $\ell$ and the sign
$\epsilon(c,\ell)=\pm$ gives the relative orientation of $c$ and $\ell$. Thus the curve $c$ is
added to the spin-network graph. The holonomies along $c$ and $\ell$ are intertwined at the new
vertex $p$ by the following contraction of the $\su(2)$ generators:
$\sum_i\tau^{(j_c)i}\tau^{(j_\ell)}_i$. This defines an intertwiner $j_c\otimes j_\ell\rightarrow
j_c\otimes j_\ell$ which can be decomposed as both representation $j_c$ and $j_\ell$ coupling to
the representation of spin 1, $((j_c\otimes \overline{\jmath_c})\arr 1 \arr(j_\ell\otimes
\overline{\jmath_\ell})$. In the usual basis of $\SU(2)$ representations with basis vectors $|j,m\rangle$, we have explicitly:
\begin{equation} \begin{split}
\sum_i(\tau_i)^{(j_1)}_{m_1 n_1}(\tau_i)^{(j_2)}_{m_2 n_2}=
& -\frac{1}{4}\sqrt{j_1(j_1+1)-n_1(n_1+1)} \sqrt{j_2(j_2+1)-n_2(n_2-1)}\ \delta_{n_1,m_1-1}\delta_{n_2,m_2+1} \\
&-\frac{1}{4} \sqrt{j_1(j_1+1)-n_1(n_1-1)} \sqrt{j_2(j_2+1)-n_2(n_2+1)}\ \delta_{n_1,m_1+1}\delta_{n_2,m_2-1} \\
&-\frac{1}{4}m_1 m_2\ \delta_{m_1, n_1}\delta_{m_2, n_2}
\end{split} \end{equation}
\begin{figure}[t] \begin{center}
\includegraphics[width=5cm]{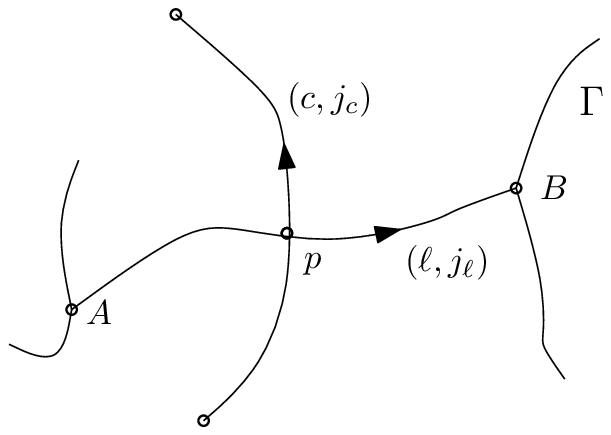}\qquad
{}\qquad\qquad
\includegraphics[width=5cm]{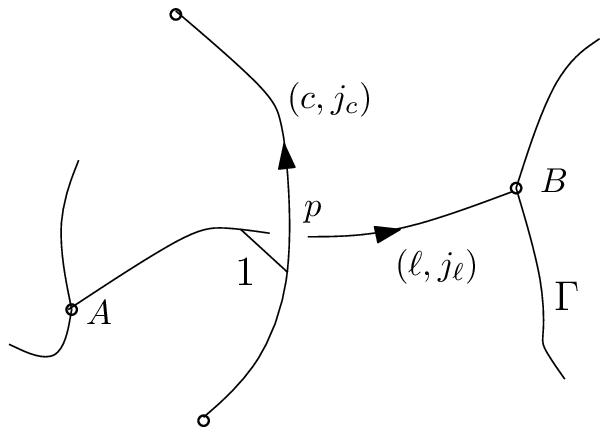}
\caption{ \label{O(c) operator}
The operator $\hat{O}(c)$ act on a spin-network graph $\Gamma$ at the intersection point $p$. In
the $\gamma\arr\infty$ limit, it creates an intertwiner at $p$ between the curves $c$ and $\ell$
represented by a new (fictious) link labeled by the spin-1 representation, explicitly
$-\frac{i}{2}(\tau^{(j_c)k})^a_b\ (\tau^{(j_\ell)}_k)^\alpha_\beta$. For finite values of $\gamma$,
we get a correction term to the intertwiner in $\gamma^{-1}$ corresponding to the trivial
intertwiner (with a spin-0 labeling the new link), explicitly $-\frac{i}{2}(\tau^{(j_c)k})^a_b\
(\tau^{(j_\ell)}_k)^\alpha_\beta+\frac{1}{2\gamma}\delta^a_b\delta^\alpha_\beta$.}
\end{center}
\end{figure}
Let us now move to finite values of $\gamma$. The Immirzi parameter deforms the quantization map:
\begin{align} \label{def operateur O}
\hat{O}(c)^\alpha_\beta &=-i\{O(c)^\alpha_\beta\ ,\,\cdot \}+\frac{1}{2\gamma}\ V(c)^\alpha_\beta\times \\  \label{def operateur Odagger}
\hat{O}^{\dagger}(c)^\alpha_\beta &=-i\{O^{\dagger}(c)^\alpha_\beta\ ,\,\cdot \}+\frac{1}{2\gamma}\
U(c)^\alpha_\beta\times
\end{align}
This reproduces the algebra generated by the triad components. For instance, let us consider the
bracket $\{O(c),O(c')\}$, with $\{p\}=c\cap c'$ being the only intersection point, $c$ going from
$x_0$ to $x_1$, and $c'$ from $y_0$ to $y_1$. The term proportional to $\gamma^{-1}$ is:
\begin{equation}
\int_c dx^a \int_{c'}dy^b \big(V^x_{x_0}\ \tau_i\ V^{x_1}_x\big)^\alpha_\beta\ \big(V^y_{y_0}\ \tau_j\ V^{y_1}_y\big)^\mu_\nu\ \{ e^i_a(x),e^j_b(y)\}
=-\frac{1}{2\gamma}\ \epsilon(c,c')\ \big(V^p_{x_0}\ \tau_i\ V^{x_1}_p\big)^\alpha_\beta\ \big(V^p_{y_0}\ \tau^i\ V^{y_1}_p\big)^\mu_\nu\
\end{equation}
We compare the commutator $[\hat{O}(c),\hat{O}(c')]$ computed from \eqref{def operateur O} to the
Poisson brackets of $O(c)$ with the holonomy $V(c')$ shown in equation \eqref{crochet O},
and we easily check that the term proportional to $\gamma^{-1}$ in $[\hat{O}(c),\hat{O}(c')]$ is
indeed:  $-i/(2\gamma)\ (\{O(c),V(c')\}-\{O(c'),V(c)\})$.
When $c$ and $c'$ do not intersect, $[\hat{O}(c),\hat{O}(c')]$ of course vanishes and there is no
contribution from the correction terms in $\gamma^{-1}$.

Finally, we can view the quantum operators $\hat{O}(c)$ and $\hat{O}^\dagger(c)$ as acting on spin
network functionals the same way as in the infinite $\gamma$ limit, but with a modified intertwiner
(see fig.\ref{O(c) operator}) taking into account the corrective holonomy term in the quantization
map, $-\frac{i}{2}\sum_k(\tau^{(j_c)k})^a_b\,
(\tau^{(j_\ell)}_k)^\alpha_\beta+\frac{1}{2\gamma}\delta^a_b\delta^\alpha_\beta$.

\subsection{Length Spectrum}

We now turn to the spectrum of lengths. For infinite $\gamma$ it is given by the square root of the
Casimir operator evaluated on the representation labeling the considered link $\ell$, that is
$\sqrt{j_\ell(j_\ell+1)}/2$. For finite $\gamma$, we introduce the following operator:
\begin{equation}
\hat{S}_j(c)=\mathrm{tr}_{(j)}\big(\hat{O}(c)\ \hat{O}^\dagger(c)\big)=
-\mathrm{tr}_{(j)}\big(\int_c V^y_{x_0}\ e_{|y}\ V^{x_1}_y\ \int_c U^{y'}_{x_1}\ e_{|y'}\ U^{x_0}_{y'}\big),
\end{equation}
for a curve $c$ going from $x_0$ to $x_1$ and colored by the spin $j$. This operator is obviously
$\SU(2)$ gauge invariant and its action is  diagonalized by spin-network states:
\begin{equation} \label{pre longueur}
\hat{S}_j(c)|\Psi\rangle = \frac{1}{4}\Big(C(j)\ j_\ell (j_\ell+1)-\frac{d_j}{\gamma^2}\Big)\ |\Psi\rangle,
\end{equation}
for a spin-network state $|\Psi\rangle$ and the curve $c$ crossing the link $\ell$ just once. The
coefficient $C(j)$ is the normalisation constant of the Killing metric in the spin $j$
representation: $\tr_{(j)}(\tau_i\tau_k)=C(j)\ \delta_{ik}$ with $C(j)=-j(j+1)d_j/3$ and
$d_j=2j+1$.

Let us see how the classical quantity $S_j(c)$ is related to the length $L(c)$ of the curve $c$.
Assume that $c$ is small i.e. of order $\epsilon\arr0^+$ in coordinates. Then we have $\int_c
V^y_{x_0}\ e_{|y}\ V^{x_1}_y \approx \epsilon\ V^q_{x_0}\ e_{|q}\ V^{x_1}_q$, where $q$ is an
arbitrary point on $c$. To lowest order in $\epsilon$, the holonomies can be approximated by the
identity. Hence, $S_j(c)\approx C(j)\delta_{ik}\int_c e^i\int_c e^k$. Consider now a generic curve
$c$. We follow the standard procedure in loop quantum gravity and cut it into $N$ small pieces
$(c_n)$. The classical length (for a smooth curve $c$) is simply:
\begin{equation} \label{longueur fonction O}
L(c)=\lim_{N\rightarrow +\infty}\sum_{n=1}^N
\sqrt{\lvert \delta_{ik}\int_{c_n} e^i\int_{c_n} e^k\rvert}
=\lim_{N\rightarrow +\infty}\sum_{n=1}^N \sqrt{\lvert C(j)^{-1}\ S_j(c_n)\rvert}
\end{equation}
Since the operator $C(j)^{-1}\hat{S}_j(c_n)$ only receives contributions from intersection points,
the limit of \eqref{pre longueur} when $c_n$ becomes very small is fairly simple: the eigenvalue is
zero if there is no intersection and is otherwise given by the r.h.s. of \eqref{pre longueur}. The
sum in \eqref{longueur fonction O} therefore contains a finite number of terms at the quantum
level, one for each intersection of $c$ with the graph of the spin-network state. For a single
intersection, we get the following spectrum:
\begin{equation}
\mathrm{Sp}(\hat{L}_j(c))=\Big\{ \frac{1}{2}\ \sqrt{j_\ell(j_\ell+1)+\frac{3}{\gamma^2j(j+1)}}\ ,\
j_\ell\in\mathbb{N^*}/2\Big\}.
\end{equation}
The effect of the Immirzi parameter is a simple shift of the usual spectrum. Therefore variations
of the length given by differences between eigenvalues remain unaffected by $\gamma$. The only
physical effect is that we will always have a non-zero minimal length in $1/\gamma\sqrt{j(j+1)}$ as
soon as we measure a distance.

Furthermore, we see that the Immirzi parameter introduces a new ambiguity: classically both the
length and the observable $C(j)^{-1}S_j(c)$ are naturally independent of the spin $j$ labeling the
curve $c$, however this spin $j$ does not drop out of the  eigenvalues of $C(j)^{-1}\hat{S}_j(c)$.
We thus obtain a set of quantum length operators $(\hat{L}_j(c))_j$ labeled by the choice of a
$\SU(2)$ representation. This phenomenon has obviously no classical equivalent and we did not find
any physical arguments allowing to fix $j$. This might be a mathematical ambiguity due to our
choice of regularization. Or it might be a physical effect
and $j$ may depend on the observer or on the matter/particles used to define the end points of the
curve $c$. Finally, we point out that the length shift goes in $1/j$ and vanishes in the limit
$j\arr\infty$.


It is possible to carry out the same analysis with a non-zero cosmological constant. As previously
mentioned, we can find $\alpha$ such that $\omega+\alpha e$, which is a $\mathfrak{su}(2)$
connection on $\Sigma$, has vanishing Poisson brackets between its components. We use this
connection to build the spin-network functionals. Then the length operator is obtained using the
operators $O(c)$ and $O^\dagger(c)$ as above. Keeping track of the coefficients coming from the
modified Poisson brackets, the eigenvalues become:
$\frac{1}{2}\lvert\frac{\gamma^2}{\gamma^2-s}\rvert^{1/2}\
\sqrt{ j_\ell(j_\ell+1)+\frac{3}{\lvert\Lambda\rvert\gamma^2j(j+1)}}$. The previous spectrum is
thus scaled by $\sqrt{\lvert\frac{\gamma^2}{\gamma^2-s}\rvert}$ while the cosmological constant
only appears in the $j$-dependent length shift proportional to $\gamma^{-2}$. Note that increasing
the absolute value of $\Lambda$ makes the $j$-dependent shift vanish. This can be understood
directly from the bracket \eqref{ee bracket}: since $\{e,e\}\propto 1/\sqrt{\lvert\Lambda\rvert}$,
we recover a commutative triad in the limit $\Lambda\arr\infty$.

\section{\label{comparison} Extension to 4d loop gravity}

The introduction of the Immirzi parameter is formally identical in our 3d setting with a
cosmological constant than in 4d: we use Hodge duality in $\mathfrak{so}(4)$ (resp. $\so(3,1)$) to
define the second bilinear form and add a new term in the action which does not change the
equations of motion. More precisely, the variables in Riemannian (resp. Lorentzian) 4d gravity are
a $\mathfrak{so}(4)$-connection (resp. $\so(3,1)$) $A^{IJ}$ ($I,J=1..4$) and a tetrad one-form
$e^I$. We define the bivector fields $E^{IJ}=\frac{1}{2}\epsilon^{IJ}_{\phantom{IJ}KL}\ e^K\wedge
e^L$. The existence of two bilinear invariant forms then enables us to build the 4d Holst action as
we did for 3d gravity in \eqref{second action}:
\begin{align}
S_H(e,A) &=2\int_M E^{IJ} \wedge \big( F+\frac{1}{\gamma}\star F\big)^{IJ} \\
  &=\int_M \epsilon_{IJKL}\ e^I \wedge e^J \wedge F^{KL}+\frac{2}{\gamma}\int_M e_I\wedge e_J\wedge F^{IJ}, \label{immirzi term}
\end{align}
with $F$  the curvature of the connection $A$. Notice that unlike in 3d \eqref{second action bis}
with non-zero $\Lambda$, the second term in \eqref{immirzi term} alone does not give general
relativity. Otherwise, similarly to eq.\eqref{self-dual decomposition} for the 3d theory, the
Immirzi parameter controls the relative contributions from the self-dual and anti-self-dual halves
of the action. Indeed, let us project the fields $E$ and $A$ on their (anti-)self-dual parts with
the projectors $(\mathrm{id}\pm\star)$. In particular, we have $F(A_\pm)=F_\pm (A)$. We get:
\begin{equation}
S_H(e,A)=\big(1+\frac{1}{\gamma}\big)\ \int_M E_{+IJ}\wedge F_+^{IJ}\ +\
\big(1-\frac{1}{\gamma}\big)\ \int_M E_{-IJ}\wedge F_-^{IJ}.
\end{equation}
Working with the bivectors $E^{IJ}$ instead of the tetrad requires us to introduce the simplicity
constraints ensuring that $E$ comes from a tetrad (see e.g. \cite{freidel depietri,clqg}).
Alternatively one can use the bivector field $\tl{E}^{IJ}$ equal to $\f12\epsilon^{IJ}{}_{KL}\,
e^K\wedge e^L+\gamma^{-1}e^I\wedge e^J$ \cite{prieto,livine oriti}. The Immirzi parameter is then
absorbed in the (simpplicity) constraints turning the topological BF theory into general
relativity:
\begin{equation}
S_H(\tilde{E},A)=\int \tl{E}^{IJ}\wedge F_{IJ}-\frac{1}{2}\phi_{IJKL}\tl{E}^{IJ}\wedge
\tl{E}^{KL}+\mu H,
\end{equation}
where $\phi$ and $\mu$ are Lagrange multipliers such that
$\phi_{IJKL}=-\phi_{JIKL}=-\phi_{IJLK}=\phi_{KLIJ}$, and
$H=a_1\phi_{IJ}^{\phantom{IJ}IJ}+a_2\phi_{IJKL}\epsilon^{IJKL}$. The parameters $a_1$ and $a_2$ are
directly related to the Immirzi parameter: $a_2/a_1=\frac{1}{4}(\gamma+\gamma^{-1})$. Such
constraints do not have any equivalent in 3d for the theory is topological.

In 4d, the parameter $\gamma$ is understood to control the amplitude of the fluctuations of the
torsion in the path integral. Indeed the Immirzi term is simply the square of the torsion up to
boundary terms\footnotemark: $\int e^I\wedge e^J\wedge F_{IJ}=\int T^I\wedge T_I$ where $T^I=d_A
e^I$ is the torsion. The relations between the Immirzi parameter and torsion have been thoroughly
studied through the coupling of the 4d theory to fermions \cite{coupling to fermions}. As we
explain in appendix, the fermion coupling through the Immirzi parameter is different in our 3d
setting than in the standard 4d analysis.


\footnotetext{
The difference of these two terms actually define the Nieh-Yan topological class defined as $\int
d_A e^I\w d_A e_I-e_I\wedge e_J\wedge F^{IJ}(A)$. }

In the canonical framework, the Immirzi parameter induces a modification of the symplectic
structure with a non-commutative connection \cite{alexandrov}.
Indeed, the simplicity constraints are second class and lead to Dirac brackets for which the triad
is commutative but for which the components of the connection do not commute with each other
anymore\footnotemark.
\footnotetext{
The usual brackets of Loop Quantum Gravity result from a change of connection and a partial gauge
fixing (the temporal gauge $e^0=0$). The Lorentz symmetry reduces to a $\SU(2)$ subgroup
\cite{barros,livine alexandrov}. The symplectic structure is then very simple and the Immirzi
parameter scales the only one non-vanishing bracket:
\be
\{E^a_i(x),A^{(\gamma)j}_b(y)\}=\gamma\ \delta_i^j\delta^a_b\ \delta^{(3)}(x-y),
\ee
where $E^a_i$ is the $\su(2)$-valued triad field and $A_a^{(\gamma)i}$ the Ashtekar-Barbero
$\mathfrak{su}(2)$-connection. This straightforwardly implies a scaling of the spectra of the
geometrical operators. The areas, given like the lengths in 3d by the Casimirs of
$\mathfrak{su}(2)$, are scaled as $\gamma$ and the volumes as $\gamma^{3/2}$ (see e.g.
\cite{carlo}).}
%

Before looking for effects on the phase space of four dimensional gravity similar to those induced
by the 3d parameter $\gamma$, let us mention a direct correspondence between $\gamma$ and some
parameters of the 4d BF theory with a cosmological term. The partition function for the theory
under investigation is formally:
\be \label{part func imm 3d}
Z_{\gamma_{3d}} = \int DA\ e^{iS_{\gamma_{3d}}(A)} = \int DA_+\
e^{\f{i}{2\sqrt{\Lambda_{3d}}}(1+\gamma_{3d}^{-1})S_{CS}(A_+)} \times \int DA_-\
e^{-\f{i}{2\sqrt{\Lambda_{3d}}}(1-\gamma_{3d}^{-1})S_{CS}(A_-)}
\ee
where we have denoted the 3d Immirzi parameter $\gamma_{3d}$ and the 3d cosmological constant
$\Lambda_{3d}$, here taken to be positive, to distinguish them from their 4d equivalents. To relate
this partition function to another in 4d, we think about Chern-Simons theory in a four dimensional
context. Indeed, given a $\SO(4)$-bundle over a 4-manifold $M$, the integral of the second Chern
form, that is the Pontryagin class, equals the Chern-Simons action on the boundary $\pp M$.
However, these quantities correspond to using the invariant non-degenerate bilinear form $(.,.)$,
equivalent to the usual trace. Using instead the second bilinear form over $\so(4)$, $\la.,.\ra$,
yields the Euler class, which turns out to be similarly related to the Chern-Simons action for
$\la.,.\ra$. Indeed:
\be
F_{IJ}(A)\wedge F^{IJ}(A) = d\Omega_{CS}^{\la,\ra}(A_{\pp M}) \qquad\text{and}\qquad
F_{IJ}(A)\wedge \big(\star F\big)^{IJ}(A) = d\Omega_{CS}^{(,)}(A_{\pp M})
\ee
where $\Omega_{CS}^{\la,\ra}$ and $\Omega_{CS}^{(,)}$ are respectively the Chern-Simons 3-forms
appearing in the actions $S$, \eqref{reformulation witten}, and $\tl{S}$, \eqref{second action}.
The partition function \eqref{part func imm 3d} can thus be interpreted in a four dimensional
context as living on the boundary of the manifold:
\be \label{part func topo class}
Z_{\gamma_{3d}} = \int DA_{\pp M}\ e^{iS_{\gamma_{3d}}(A_{\pp M})} = \int DA_+\
e^{\f{i}{2\sqrt{\Lambda_{3d}}}(1+\gamma_{3d}^{-1})\int_M F_{+IJ}\wedge F_+^{IJ}} \times \int DA_-\
e^{-\f{i}{2\sqrt{\Lambda_{3d}}}(1-\gamma_{3d}^{-1})\int_M F_{-IJ}\wedge F_-^{IJ}}
\ee

The second key observation is that these classes, which are bundle invariants, can be obtained in a
first order formulation from the 4d BF theory with a cosmological term. Indeed, consider the
following action:
\be
S_{\gamma,\Lambda}(E,A) = \int_M E_{IJ}\wedge\big(F+\f{1}{\gamma}\star F\big)^{IJ} -\f{\Lambda}{2}
E_{IJ}\wedge (\star E)^{IJ}
\ee
Notice that restricting $E$ to be of the form $E=\star(e\wedge e)$ for a tetrad field $e$
reproduces the Holst action $S_H(e,A)$ supplemented with a cosmological constant $\Lambda$. The
partition function for $S_{\gamma,\Lambda}$ can be easily worked out using its
self-dual/anti-self-dual decomposition:
\be
S_{\gamma,\Lambda}(B,A) = \int_M B_{+IJ}\wedge F_+^{IJ} - \f{\Lambda}{2\big(1+\gamma^{-1}\big)^2}
B_{+IJ}\wedge B_+^{IJ} -\int_M B_{-IJ}\wedge F_-^{IJ} - \f{\Lambda}{2\big(1-\gamma^{-1}\big)^2}
B_{-IJ}\wedge B_-^{IJ}
\ee
in which we have rescaled the field $E$: $B_{\pm} \equiv (\gamma^{-1}\pm 1) E_\pm$. Formally
performing the Gaussian integral over the field $B$, we are lead to the following action, which is
a superposition of the Pontryagin and Euler classes:
\be
S_{\gamma,\Lambda}(A) = \f{(1+\gamma^{-1})^2}{2\Lambda} \int_M F_{+IJ}\wedge F_+^{IJ} -
\f{(1-\gamma^{-1})^2}{2\Lambda} \int_M F_{-IJ}\wedge F_-^{IJ}
\ee
As in the 3d case, the coupling constants of the self-dual and anti-self-dual parts do not simply
defer by a sign because of the presence of the Immirzi parameter. Moreover, the interpretation of
the action $S_{\gamma_{3d}}$ in terms of 4d topological classes enables to identify the coupling
constants between the 3d and 4d cases, by comparing the partition function for $S_{\gamma,\Lambda}$
with
\eqref{part func topo class}. This leads to: $\gamma_{3d} = \f{1}{2}\big(\gamma+\gamma^{-1}\big)$,
and $\sqrt{\Lambda_{3d}} = \Lambda\gamma^2/(1+\gamma^2)$.

Another interesting correspondence gives simple relations between the parameters. Consider the 4d
BF action, but instead of the term $E\wedge \star F$ previously used, use the Hodge duality to
introduce a new term quadratic in the field $E$:
\be
S_{\beta,\Lambda}(E,A) = \int_M E_{IJ}\wedge F^{IJ} -\f{\Lambda}{2}E_{IJ}\wedge\Big(\star
E+\f{1}{\beta} E\Big)^{IJ}
\ee
Notice however that the added term vanishes when evaluating it on the specific configurations
$E=\star(e\wedge e)$. After integration over the field $E$ into the path integral, we are again
lead to the Pontryagin and Euler classes, but the coupling constants for the self-dual and
anti-self-dual parts are now different functions of the parameters:
\be \label{second comp}
S_{\beta,\Lambda}(A) = \f{1}{2\Lambda\big(1+\beta^{-1}\big)} \int_M F_{+IJ}\wedge F_+^{IJ} +
\f{1}{2\Lambda\big(\beta^{-1}-1\big)} \int_M F_{-IJ}\wedge F_-^{IJ}
\ee
Comparing \eqref{second comp} and \eqref{part func topo class} gives the relations: $\gamma_{3d} =
-\beta$ and $\sqrt{\Lambda_{3d}} = \Lambda\big(1-\f{1}{\gamma^2}\big)$.

The 3d symplectic struture which we studied here can be extended to 4d by adding these bundle
invariants to the 4d gravity action. In particular, instead of only considering the torsion
squared, we should also consider terms involving the curvature squared. To this purpose, let us
introduce the Pontryagin class and the Euler class into the 4d BF action:
\begin{gather} \label{BF+ topo classes}
S(A,E)=\f{1}{2}\int_M E^{IJ}\wedge F_{IJ}+\theta_1\ g_{IJKL}\ F^{IJ}\wedge F^{KL}, \\
\textrm{with}\quad g_{IJKL}=\f{1}{2}\big(\delta_{IK}\delta_{JL}-\delta_{IL}\delta_{JK}\big)+\f{\theta_2}{2\theta_1}\
\eps_{IJKL}.
\end{gather}
$g_{IJKL}$ defines a metric on the Lie algebra $\so(4)$. The Pontryagin and Euler classes do not
depend on the connection (because of the Bianchi identity) and thus do not modify the equations of
motion neither for BF theory nor for gravity (the Pontryagin class is the $\theta$-term of
Yang-Mills theories). $\theta_1$ and $\theta_2$ control the relative contributions of the self-dual
and anti-self-dual parts of the action :
\be \label{(anti)self-dual contributions + topo terms}
S(A,E)=\int_M\f{1}{2}\ E_+^{IJ}\wedge F_{+IJ}+(\theta_1+\theta_2)\ F_+^{IJ}\wedge
F_{+IJ}+\int_M\f{1}{2}\ E_-^{IJ}\wedge F_{-IJ}+(\theta_1-\theta_2)\ F_-^{IJ}\wedge F_{-IJ}.
\ee
The canonical analysis of \eqref{BF+ topo classes} is straightforward (see \cite{montesinos} for a
detailed analysis) and is very similar to the 3d case that we studied in the previous sections: the
momentum conjugated to the connection is $\Pi^a_{IJ}=E^a_{IJ}+2\theta_1 B^a_{IJ}$, with $E^a_{IJ} =
\eps^{abc}\ E_{bcIJ}$ and $B^a_{IJ}=\eps^{abc}g_{IJKL}F^{KL}_{bc}$. This yields a non-commuting $E$
triad field:
\be \label{E bracket topo terms in 4d}
\Big\{ E^{a}_{IJ}(x), E^b_{KL}(y)\Big\} = 4\theta_1\Big\{ \eps^{abc}\  g_{IJ}^{\phantom{IJ}MN}\ D_c^{(x)} \delta^{KL}_{MN}\ \delta^{(3)}(x-y)
-\big((IJ)\leftrightarrow (KL),a\leftrightarrow b,x\leftrightarrow y\big)\Big\},
\ee
with $\delta^{KL}_{MN} = \f{1}{2}(\delta^K_M\delta^L_N-\delta^K_N\delta^L_M)$. In this equation,
the covariant derivative $D_c$ is taken to act on the upper indices of $\delta^{KL}_{MN}$. This
bracket is thus proportional to the covariant derivative while in 3d, it is proportional to the
identity \eqref{triad brackets}. This is related to the fact that the momenta are shifted by the
connection in 3d and by the curvature in 4d. As far as the phase space is concerned, notice a
particular duality in 4d: while the torsion squared term $T^I\wedge T_I$ takes part in the
non-commutativity of the connection, curvature squared terms $g_{IJKL}F^{IJ}\wedge F^{KL}$ are
responsible for the non-commutativity of the triad field.

We now turn to GR adding these topological terms to the Palatini action and proceeding to the
ususal canonical analysis (following \cite{barros,ashtekar's book}). Before imposing the
second-class constraints, the (unreduced) phase space is that of BF theory. The Hamiltonian is
however different, made of the Gauss, diffeomorphism and scalar constraints. One can easily check
that the constraint algebra is not modified by the addition of the topological terms: any smearing
of the bracket \eqref{E bracket topo terms in 4d} over $\Sigma\times\Sigma$ identically vanishes.

Following \cite{barros}, the second-class constraints are solved by writing $E^a_{IJ}$ as
$E^a_{IJ}=\f{1}{2}(n_I E^a_J-n_J E^a_I)$ with a timelike unit vector $n^I$. The standard LQG
approach relies on gauge fixing the time-like direction $n$ with the choice $n^I\equiv\,(1,0,0,0)$,
which makes $E^a_{IJ}$ a pure boost. The canonical variables are then the triad $E^a_i=E^a_{0i}$
and the extrinsic curvature $K_a^i=A_a^{0i}$. One should also solve the boost components of the
Gauss constraint, which states that the rotational components of the connection form the $\SU(2)$
spin-connection compatible with the triad $E^a_i$, that is $A_a^{ij}=-\eps^{ij}_{\phantom{ij}k}\
\Gamma^k_a(E)$. However, in the presence of the topological terms, the canonical momenta
$\Pi_{IJ}^a$ acquire a non-zero rotational part, let's call it $B^a_{ij}$. These components become
functions $B^a_{ij}(E)$ of the triad after gauge fixing. Thus the canonical momenta of $K_a^i$ and
$E^a_i$ have to be extracted from the following kinetic terms of the action (we have set
$\theta_2=0$ for simplicity) :
\be \label{topo terms for GR}
S_{\mathrm{kin},\theta_1} = \int d^4x\ \Big(E^a_i+4\theta_1\eps^{abc}\nabla_b K_{ci}\Big)\pp_t
K_a^i+2\theta_1\eps^{abc}\ \Big(R_{bci}+\eps_{ijk}\ K_b^j K_c^j\Big) \pp_t\Gamma_a^i,
\ee
$\nabla$ and $R$ being respectively the covariant derivative operator and the curvature of the
spin-connection $\Gamma(E)$. However, this result takes far from the 3d situation studied here and
from the usual context of LQG.

We can nevertheless notice that the situation gets much simpler, and indeed very close to the 3d
case, when looking at the self-dual formulation of gravity. In this case, we set the couplings
$\gamma=1$ and $\theta_1=\theta_2$. The theory is then formulated in terms of $\SU(2)$ variables
right from the start (without any gauge fixing). Moreover, there is no additional second-class
constraints to the Hamiltonian, the momentum $\Pi^a_{+i}$ being an arbitrary self-dual field. Thus
we only need to consider the self-dual terms of the action \eqref{(anti)self-dual contributions +
topo terms}. We use the following notation, for all self-dual fields, $X^i=X_+^{0i}$, dropping the
indice + to emphasize the fact that all references to the anti-self-dual sector disappear. The
phase space is now parametrized by pairs of canonically conjugate variables consisting of the
connection $A_a^i$ and a triad shifted by the curvature of $A$, $\Pi_i^a=E^a_i+2\theta\eps^{abc}\
F_{bci}$, with $\theta=2\theta_1$. The resulting brackets are the same as those of BF theory with
topological terms, for the group $\SU(2)$:
\begin{align}
\{A_a^i(x),A_b^j(y)\} &= 0 \\
\{A_a^i(x),E^b_j(y)\} &= \delta_a^b\delta^i_j\ \delta^{(3)}(x-y) \\
\{E^a_i(x),E^b_j(y)\} &=
4\theta\ \Big[ \eps^{abc}\ D_c^{(x)}\delta_{ij}\delta^{(3)}(x-y) - \Big(a\leftrightarrow
b,x\leftrightarrow y,i\leftrightarrow j\Big)\Big]
\end{align}
with $D_c\delta_{ij} = \pp_c\delta_{ij}+\eps_{ikj}\ A_c^k$. Notice that this is exactly the
situation described in \cite{alejandro danilo}, but only for an Immirzi parameter fixed to
$\gamma=1$ (or similarly $\gamma=-1$). As shown in \cite{alejandro danilo}, the usual flux
variables of LQG are undefined for such a canonical structure. Nevertheless, given the similarity
of the phase space with the 3d structures that we studied, we propose an alternative strategy which
could be fruitful to solve this issue: quantize another algebra considering one-insertion loops
variables instead of flux variables, so that the additional term required to satisfy the new
commutation relations at the quantum level is simply given by a holonomy.

For arbitrary values of the couplings $\gamma,\theta_1,\theta_2$, we have to deal once again with
the non-zero rotational part of $\Pi^a_{IJ}$ as above for eq.\eqref{topo terms for GR}. Indeed,
considering the action:
\be
S_{\gamma,\theta_1,\theta_2}(A,e) = \f{1}{4}\ S_H(e,A) + \theta_1 \int F^{IJ}\wedge
\big(F_{IJ}+\f{\theta_2}{2\theta_1}\ \eps_{IJKL}\ F^{KL}\big),
\ee
one finds the following kinetic terms for the triad and the extrinsic curvature:
\be
S_{\mathrm{kin},\gamma,\theta_1,\theta_2} = \int d^4x\ \f{1}{\gamma}\ E^a_i\
\pp_t\big(\Gamma_a^i-\gamma K_a^i\big) + 4\theta_2\eps^{abc}\ \nabla_b
K_{ci}\ \pp_t\big(\Gamma_a^i-\f{\theta_1}{\theta_2}\ K_a^i\big) - 2\theta_1\eps^{abc}\
\Big(R_{bci}+\eps_{ijk}\ K_b^j K_c^k\Big)\pp_t\big(\Gamma_a^i-\f{\theta_2}{\theta_1}\ K_a^i\big).
\ee
The first term of the r.h.s. is the usual one for LQG, while the second and the third are
respectively the boost and the rotational parts coming from the kinetic terms of the topological
classes. There is a special case
$\gamma=\theta_1/\theta_2$ for which we can formulate the canonical structure can be formulated in
term of the connection variables $\Gamma^\pm\,\equiv\,\Gamma-\gamma^{\pm 1} K$ as in Holst's
analysis
\cite{holst}:
\be
S_{\mathrm{kin},\gamma=\f{\theta_1}{\theta_2}} = \int d^4x\ \f{1}{\gamma}\ \Big(E^a_i +
4\theta_1\eps^{abc}\ \nabla_b K_{ci}\Big)\pp_t\big(\Gamma_a^i-\gamma K_a^i\big) -
2\theta_1\eps^{abc}\ \Big(R_{bci}+\eps_{ijk}\ K_b^j K_c^k\Big)\pp_t\big(\Gamma_a^i-\f{1}{\gamma}\
K_a^i\big).
\ee
As we see, the second kinetic term can be absorbed in a simple shift of the triad variable.
Finally, we point out that the cases $\gamma=\pm 1$ are the only choices that make it possible to
completely re-absorb the rotational components of $\Pi^a_{IJ}$ in the momenta conjugate to
$\Gamma_a^i-\gamma K_a^i$.

\section*{Conclusion}

A Immirzi parameter for three-dimensional gravity can be formally introduced the same way as it
appears in the four-dimensional Holst action for general relativity. However the Poisson brackets
become more intricate in the 3d case and one has then to deal with a non-commuting triad field. The
length spectrum can nevertheless be derived using modified flux operators. The contribution of the
Immirzi parameter is not a scaling of the geometric spectra like in 4d, but a simple constant shift
of the eigenvalues. The drawback of our approach is that we obtain a one-parameter family of length
operators labeled by a $\SU(2)$ representation which are all equivalent classically but not at the
quantum level. This ambiguity is not yet fully understood and deserves further investigation. We
can nevertheless compare with the regularization ambiguity for the quantization Hamiltonian
constraint in 3+1d loop quantum gravity (see e.g. \cite{thomaslqg}).


In the final section, we compare our 3d setting and the standard 4d Immirzi parameter. Although
they both come from the fact that there exists two bilinear forms on the Lie algebra $\sl(2,\C)$,
they turn out to be rather different effects on the phase space. While the 4d Immirzi parameter
relates to a torsion squared term in the action, our 3d Immirzi parameter is as expected better
compared to the $\theta$-parameters for the topological Pontryagin and Euler classes given as
squares of the curvature. This should be related to recent investigation on the effect of a
$\theta$ parameter in canonical loop gravity \cite{alejandro danilo}.

To conclude, we believe that the effects of this 3d Immirzi parameter on the observables of 3d BF
theory and their state sum representation should be investigated in more details. This would
establish further links between knot invariants and loop quantum gravity observables (in 3d).
Furthermore, the moot point is whether or not it is possible to write a spin foam model of the
Ponzano-Regge type (as a state sum) for 3d gravity with $\gamma\ne 0$. Indeed the levels of the
(anti-)self-dual Chern-Simons theories are now different and the overall path integral is not given
by the Turaev-Viro model anymore. Understanding how to deform the Turaev-Viro ansatz to accomodate
such an extra term in 3d would provide a state sum representation of $\SU(2)$ Chern-Simons theory
but would also help understanding how to deal with similar deformations in 4d gravity.


\section*{Acknowledgements}

We would like to thank Karim Noui for many discussions on 3d quantum gravity.

\appendix
\section{\label{couplage fermions} Coupling to Fermions}

We have seen that the parameter $\gamma$ labels a family of classically equivalent theories
describing pure gravity. However, this is not true anymore in the presence of fermions because they
are a source of torsion. In 4d like in 3d, the metric formulation uses the Levi-Civita connection
which is torsion-free. In the Palatini first order framework with independent connection and
vierbein fields, the vanishing torsion is implemented by the equation $d_\omega e=0$. In 4d,
fermions introduce torsion in the theory and the new equation of motion is of the form ``$d_\omega
e=$ fermionic current" \cite{coupling to fermions} which explicitly involves the Immirzi parameter
$\gamma$. Then it appears in the effective action for fermions as a coupling constant for a
4-fermions interaction (Einstein-Cartan term):
\begin{equation} \label{int einstein cartan}
S_{\mathrm{int}}(e,\psi)=-\frac{3}{2}\pi G \frac{\gamma^2}{\gamma^2+1}\int_M d^4x \sqrt{g}\
\big(\overline{\psi}\gamma_5\gamma_I\psi\big)\ \big(\overline{\psi}\gamma_5\gamma^I\psi\big).
\end{equation}
In 3d, the parameter $\gamma$ that we introduced is still related to torsion, although it plays a
rather different role. Consider the following action for gravity at $\Lambda=0$, restoring the
Newton constant $G$:
\be
S_\gamma(e,\omega) = \f{2}{G} \int_M e^i\wedge F_i[\omega] + \frac{1}{\gamma}\int_M \omega^i\wedge
d\omega_i+\frac{1}{3}\epsilon_{ijk} \omega^i\wedge \omega^j\wedge\omega^k.
\ee
Fermions transform under the fundamental representation of $\SU(2)$ (there is neither Weyl spinors,
nor axial and vector current in 3d). They are coupled to gravity through the following minimal
coupling interaction:
\begin{equation}
S_F(e,\omega,\psi,\psi^\dagger)=\frac{i}{2}\int d^3x\,(\det e)\, \big(\psi^\dagger\sigma^i e_i^\mu
D_\mu\psi-(D_\mu\psi)^\dagger\sigma^ie_i^\mu \psi\big)
\end{equation}
where $\det e$ is the determinant of the triad and $D=d_\om$, or explicitly
$D_\mu=\partial_\mu-(i/2)\, \omega_\mu^i\sigma_i$, is the covariant derivation for $\omega$. For
$\gamma\arr \infty$, the equation of motion for the torsion is:
\be
\epsilon^{\mu\nu\lambda}\ D_\nu e_\lambda^i=-\frac{G}{4}(\det e)\, e^{\mu i}\psi^\dagger\psi,
\ee
which shows explicitly that fermions are a source of torsion in the theory. This equation is solved
for the connection by writing $\omega$ as the sum of a torsion-free part $\Gamma[e]$, determined by
$de+[\Gamma[e],e]=0$, and a part containing the torsion: $\omega^i=\Gamma[e]^i+C^i$. Then the
action can be expressed in terms of the triad and the fermionic fields. Here we get:
\be
C_\mu^i=-\frac{G}{8}\ e_\mu^i\ \psi^\dagger\psi,
\ee
with a torsion of order $G$, like for 4d gravity. Inserting the expression of
$\omega[e,\psi,\psi^\dagger]$ into the action, we recover general relativity with fermions in the
second-order formalism with an additional interaction term:
\be
S(e,\psi,\psi^\dagger)=\frac{1}{G}\int d^3x\ \epsilon^{\mu\nu\lambda}\delta_{ij}\ e_\mu^i\
F_{\nu\lambda}^j\big[\Gamma[e]\big]+\frac{i}{2}\int d^3x(e)\Big(\psi^\dagger\sigma^i e_i^\mu
\nabla_\mu\psi-(\nabla_\mu\psi)^\dagger\sigma^ie_i^\mu \psi\Big) - \f{3}{32}\ G\int d^3x(e)\
\big(\psi^{\dagger}\psi\big)^2,
\ee
where $\nabla_\mu=\partial_\mu-(i/2)\ \Gamma^i_\mu[e]\sigma_i$.

For 4d gravity, the first-order and second-order formalisms are not equivalent. They differ from
each other due to the 4-fermion interaction \eqref{int einstein cartan} which is proportional to
the Newton constant $G$. This term remains even in the with $\gamma\arr\infty$ limit. The situation
is similar in 3d: torsion is responsible for a 4-fermion coupling of order $G$. This quartic
interaction term is simply the squared fermionic density.

The situation becomes  more intricate for finite values of $\gamma$. First the torsion equation of
motion acquires a curvature term proportional to $G\gamma^{-1}$:
\be \label{equation torsion immirzi}
\epsilon^{\mu\nu\lambda}\Big(D_\nu e_\lambda^i+\frac{G}{2\gamma}F_{\nu\lambda}^i[\omega]\Big)
=-\frac{G}{4} \,\det e\, e^{\mu i}\psi^\dagger\psi.
\ee
However the curvature is also non-zero in presence of fermions independently of the parameter
$\gamma$. Varying the action with respect to the triad, we indeed have:
\be
\epsilon^{\mu\nu\lambda}\ F_{\nu\lambda}^i[\omega] = -\frac{iG}{2}\epsilon^{\mu\nu\lambda}\ \epsilon^{i}_{\phantom{i}jk}\ e_\nu^j\ \Big(\psi^\dagger\sigma^k D_\lambda\psi-(D_\lambda\psi)^\dagger\sigma^k\psi\Big)
\ee
Inserting this expression of $F[\omega]$ into \eqref{equation torsion immirzi}, we extract the
correction to the connection $\om$ due to the torsion:
\be \label{contorsion}
C_\mu^i=\frac{1}{1-\frac{G^2}{4\gamma}\psi^\dagger\psi}\ \Big(i\f{G^2}{4\gamma}\
\big(\psi^{\dagger}\sigma^i\nabla_\mu\psi-(\nabla_\mu\psi)^{\dagger}\sigma^i\psi\big)-\f{G}{8}\
e_\mu^i\ \psi^{\dagger}\psi \Big).
\ee
Putting this equation back into the action, one obtains an action which is not polynomial into the
fermionic fields anymore. In particular, the denominator of $C_\mu^i$ imposes a limit to the
fermionic density: $4|\psi\dag\psi|\le G^2\gamma^{-1}$. This is an intriguing new role for this 3d
Immirzi parameter.  Assuming that $G$ very small, this denominator can be expanded in powers of
$G^2$ and leads to a polynomial expansion of the action. At order $G^2$, the denominator of
$C_\mu^i$ can actually be neglected, and the action becomes:
\be \begin{split}
S_\gamma(e,\psi,\psi^\dagger) = &\frac{1}{G}\int d^3x\ \eps^{\mu\nu\lambda}\delta_{ij}\ e_\mu^i\ F_{\nu\lambda}^j\big[\Gamma[e]\big] + S_F(e,\Gamma[e],\psi,\psi^\dagger) + \f{1}{\gamma}S_{CS}(\Gamma[e]) \\
 &- \f{3}{32}\ G\int d^3x(e)\ \big(\psi^{\dagger}\psi\big)^2 - \f{1}{8\gamma}\ G\int d^3x\ \eps^{\mu\nu\lambda}\ e_\mu^i F_{\nu\lambda i}\big[\Gamma[e]\big]\ \psi^\dagger\psi \\
 &+\f{i}{4\gamma^2}\ G^2 \int d^3x\ \eps^{\mu\nu\lambda}\ \Big(\psi^\dagger\sigma_i\nabla_\mu\psi - (\nabla_\mu\psi)^\dagger\sigma_i\psi\Big)\ F_{\nu\lambda}^i\big[\gamma[e]\big]
 +\dots
\end{split} \ee
It is thus clear that the 3d Immirzi parameter $\gamma$ generates classically non-equivalent
theories for gravity coupled to matter. Moreover it provides us with a new comparison with the 4d
Immirzi parameter, which turns out to be related to torsion in a different way. In particular, in
contrast with the 4d case, the torsion does not vanish in 3d in the limit $\gamma\arr 0$.


\end{document}